\documentclass[twocolumn,showpacs,preprintnumbers,amsmath,amssymb,floatfix,boldsymbol]{revtex4}
\usepackage{graphics} 
\usepackage{epsfig}
\usepackage{graphicx}
\usepackage{dcolumn}
\usepackage{amsbsy}

\begin{document}

\title{Band-edge diagrams for strained III-V semiconductor quantum wells, wires, and dots} 
\author{C. E. Pryor\footnote{Electronic mail: craig-pryor@uiowa.edu}}
\affiliation{ Optical Science and Technology Center and Department of Physics and Astronomy, University of Iowa, Iowa City, Iowa, 52242, USA}

\author{ M.-E. Pistol\footnote{Electronic mail: mats-erik.pistol@ftf.lth.se}}
\affiliation{Solid State Physics, Lund University,
P.O. Box 118, SE-221 00 Lund, Sweden}
\date{\today} 

\begin{abstract}
{\footnotesize
We have calculated band-edge energies for most combinations of zincblende AlN, GaN, 
InN, GaP, GaAs, InP, InAs, GaSb and InSb in which one material is strained to the other.  
Calculations were  done for three different geometries, quantum wells, wires, and dots, and
mean effective masses were computed in order to estimate confinement energies.
For quantum wells, we have also calculated  band-edges  for ternary alloys. 
Energy gaps, including confinement, may be easily and accurately estimated using band energies and a simple effective mass approximation, yielding excellent agreement with experimental results.
By calculating all material combinations we have identified novel and interesting material combinations, such as artificial donors, that have not been experimentally realized.
The calculations were perfomed using strain-dependent ${\boldsymbol {k\cdot p}}$-theory and provide a comprehensive overview of band structures for strained heterostructures. 
}
\end{abstract}

\pacs{73.23.Ad, 73.63.Rt, 85.35.Ds}
\maketitle

\section{Introduction}
Diagrams of band edges vs material composition or lattice constant  for bulk semiconductors  \cite{Tiwari.apl.1992} have proved indispensable for bandgap engineering.
For heterostructures containing materials with two different lattice constants, however,  such diagrams are problematic because the band energies are modified by strain.  
A familiar example is the bandgap of InAs ($E_{g} \approx 0.41  \, \mathrm{ eV}$) which approximately doubles when grown on GaAs.
The strain depends on the lattice mismatch, the elastic properties of both materials, and the geometry.
The energy shift in turn depends on the strain and the electronic material parameters, primarily the deformation potentials.
Theoretical treatments often rely on approximating a structure as a slab because it is amenable to simple analytic calculations, in spite of the fact that wires and dots are poorly approximated by a slab.

To address this problem, we have calculated band energies for heterostructures consisting of direct gap binary III-V compounds, including zincblende nitrides, and  including GaP and AlN as substrate materials.
We have taken the geometries of the embedded materials to be slabs, (quantum wells), circular wires, and lens shaped dots. 
We have restricted ourselves to direct gap materials because they are of the most interest for optical applications, and because deformation potentials for indirect materials are usually less well known.
For quantum wells we have also calculated energies for ternary alloys grown on substrates having lattice constants between $5.4 \, \mathrm{nm} $ and  $6.5 \, \mathrm{nm} $. 
The results provide a systematic and comprehensive resource for the design and interpretation of strained low-dimensional heterostructures.

For some material combinations the extremely large lattice mismatch makes growth of pseudomorphic structures on large area substrates problematic. 
However, there has been progress in the fabrication of heterostructures in freestanding wires  \cite{Bjork.apl.2002} with diameters between  $10 \, \mathrm{nm} $ and  $100 \, \mathrm{nm} $.  
Wires allow the realization of larger mismatches in part because of strain relaxation in the barrier material, leading to smaller strain in the well material.
The lack of misfit dislocations may allow heterostructures to be grown beyond the classical limit of Matthews and Blakeslee \cite{Matthews.jcg.1974}.  
While we do not consider such wires here, the results for slabs are applicable.
Progress may also occur in the growth of quantum dots, in particular in the growth of dots under tension.

\section{method}
The calculations were performed by a method that has been previously described \cite{Pryor.prb.1997}.
The strain was calculated using continuum elasticity and the finite element method.
Energies were then computed by taking the local value of the strain and computing the energy from the 8-band strain-dependent ${\boldsymbol {k \cdot p} }$ Hamiltonian at ${\boldsymbol k } = 0 $.  
(With  ${\boldsymbol k } = 0 $, this is in fact a 6+1 band model since the valence and conduction bands are coupled by terms proportional to $\boldsymbol k$).
All material parameters were taken from Ref.  \cite{Vurgaftman.jap.2001} with $T=0~K$, and all nitrides were taken to be in the zincblende form.
The computational grids were $100 \times 100 \times 100$  and $100 \times 100$ for dots and wires respectively.  For wells only two sites were needed since the strain is biaxial.
As a check on the results, the calculations were done independently by each author.  For wires and dots the independent calculations were done with the same software, but the quantum well calculations were checked by using different programs. 

\begin{figure}
\includegraphics[width=1.0\columnwidth]{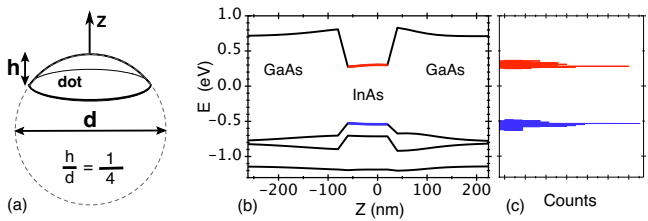}
\caption{(a) The lens-shaped geometry used for all quantum dot calculations.  The dot height is 1/4 the diameter of sphere out of which the lens is cut.
(b) Band structure as a function of position along an InAs/GaAs dot's axis of symmetry.  
The energies are calculated  for ${\mathbf k} = 0$ using the local value of the strain.
(c) Histogram of the conduction and valence band edges throughout the full volume of the dot.}
\label{schematic}
\end{figure}

For bulk materials the band edges are characterized by single numbers, as are the band edges for strained quantum wells since their strain is homogeneous.  
For wires and dots, however, the strain is inhomogenous, and hence the band edge (computed as the  ${\boldsymbol k} = 0 $ energy) varies around the structure (Fig. ~\ref{schematic}).
To represent complex strain distributions, histograms of the band edges were computed for both wires and dots, although wires have significantly more homogeneous strain than dots. 

While a band edge is well-defined for bulk materials, wires, and wells, quantum dots do not have bands.  
Nonetheless, it is useful to consider the local band edge that would be present in bulk material having the same strain as a location in the heterostructure.  
For a sufficiently large heterostructure such band edges give the potential in the effective mass approximation. 
Even for nanometer scale structures the effective potential obtained from the local band edge provide a useful estimate of electronic energies \cite{Pryor.prb.1998}.

Since the equations governing the strain are scale invariant, the strain and band energies do not vary with the size of the structure, and
the band-edge results are applicable to any size structure.
To facilitate calculation of the confinement energy, we have computed mean effective masses for the structures.
These were spatially averaged over the well material, and over directions in which there was confinement.
 \begin{figure}
\includegraphics[width=1.0\columnwidth]{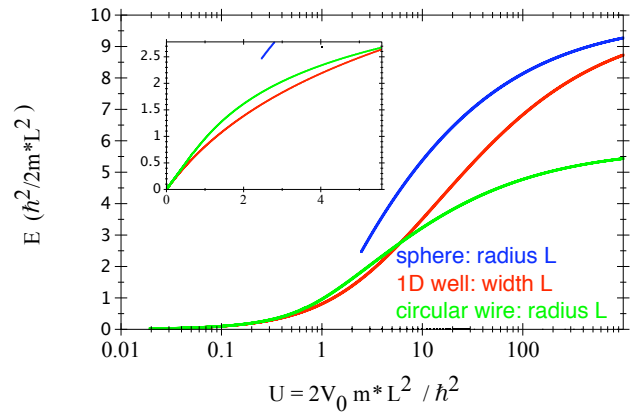}
\caption{ Confinement energy as a function of the barrier height $V_0$, the effective mass inside the well $m^*$, and the dimension of the structure, $L$.  For $V_0=1~\mathrm{eV}$, $m^*=m_e$, and $L=1~\mathrm{nm}$, $U=13.123$. There is always a bound state for the 1D and 2D cases, but for 3D a minimum potential strength is required.
}
\label{confinementEnergy}
\end{figure}
Given the dimensions of a nanostructure, the barrier height, and the effective mass one can approximate the confinement energy using a simple effective mass model.
Figure ~\ref{confinementEnergy} gives the confinement energies for quantum wells,  circular wires, and spheres with various barrier heights, effective masses, and sizes.  

Self assembled quantum dots have complex and variable shapes, but we have found that modeling them as circular cylinders works well for estimating the confinement energy.  
Assuming a dot to be a lens shaped cap with base diameter $d$ and height $h$ we model it as a circular cylinder of diameter $d$ and a height $h_{cyl}$ that gives the cylinder the same volume as the lens-shaped cap ( $V =  \pi h \left(  2 d^2 + 4 h^2\right) /24$).  
Since typically $h << d$, the confinement energy is dominated by the 1-D confinement along the shorter direction.

This undertaking would have been considerably more difficult without a recent publication of critically reviewed parameters for all binary III-V compounds and many alloys \cite{Vurgaftman.jap.2001}.   
Material parameters are subject to varying degrees of uncertainty, and the calculations presented here are primarily constrained by the accuracy of those tabulated parameters. 
The  parameters for the nitrides are especially uncertain, and mostly based on theoretical calculations.
For example, the band-gap of wurtzite InN has recently been reevaluated by a large factor   \cite{Matsuoka.apl.2002}.
Previous investigations have shown that calculations for strained heterostructures vary in their sensitivity to material parameters, with the deformation potentials and Luttinger parameters being the most important \cite{Pryor.prb.1999}.
In addition, the material parameters are known with varying degrees of accuracy. 
For example, lattice and elastic constants are known to four or more significant figures, while Luttinger parameters are typically only certain to within a few tens of percent.

\section{Results and discussion}
\label{diagrams}

We first consider dots, wires and wells consisting of combinations of binary materials.
For each substrate material we have plotted the conduction and valence  energies of the embedded material, along with the value for the unstrained bulk substrate material.
The energy of the top of the valence band for unstrained InSb has been taken as a reference level and set to zero. 
The spatial variation of the band edges in dots and wires is displayed as a histogram in which the shading of the lines is proportional to the frequency of a particular energy. 
An example of the potential profile in a quantum dot along with the corresponding histogram is shown in Fig. ~\ref{schematic}.
The range of energies is larger for dots than wires as a consequence of the larger strain inhomogeneity in dots.

\subsection{AlN, GaN, and InN substrates}

\begin{figure}
\includegraphics[width=1.0\columnwidth]{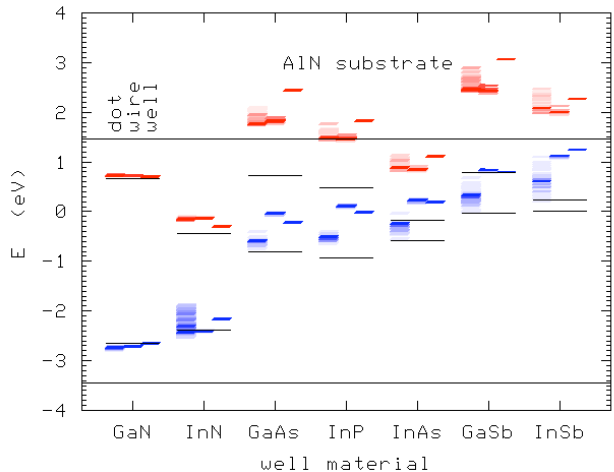}
\caption{Diagram of band energies vs composition for dot, wire, and well structures on an AlN substrate.
(All nitrides are zincblende.)
The two long lines spanning the graph indicate the conduction and valence band energies for the unstrained substrate material.  
The lines of medium length indicate the unstrained valence and conduction energies for the different well materials.
The short lines show the valence and conduction energies for dots, wires, and wells of the indicated composition with strain effects included.
For dots and wires the shading of the lines is proportional to the 
 of material with the indicated energy.
All energies are calculated  for ${\mathbf k} = 0$ using the local value of the strain.
}
\label{AlN}
\end{figure}

\begin{figure}
\includegraphics[width=1.0\columnwidth]{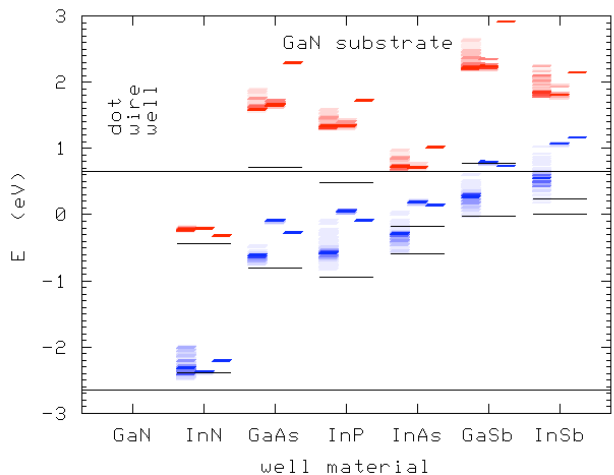}
\caption{Band edge diagram of strained dots, wires and wells on GaN.
All nitrides are zincblende.
}
\label{GaN}
\end{figure}

\begin{figure}
\includegraphics[width=1.0\columnwidth]{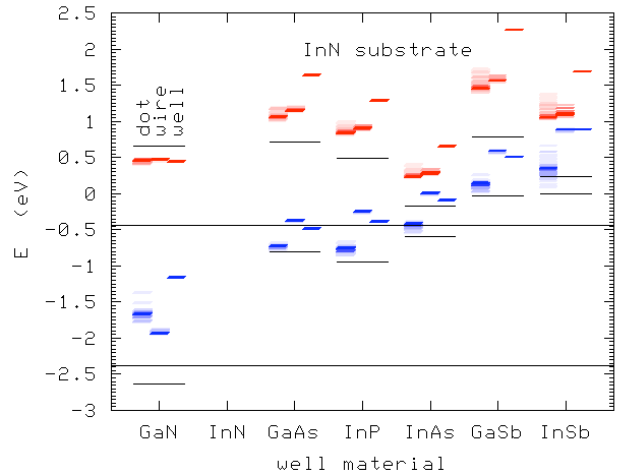}
\caption{Band edge diagram of strained dots, wires and wells on InN.}
\label{InN}
\end{figure}

Figures \ref{AlN} - \ref{InN} show the energies for  zincblende GaN, InN, GaAs, InP, InAs, GaSb, and InSb strained to AlN, GaN and InN, respectively.  
We find that for AlN substrates all of the nitrides are type I, 
while all of the non-nitrides except InAs are type II, with only hole confinement.
On GaN substrates we find only InN to be type I. 
For GaSb and InSb wires and wells there is a broken gap, with the valence maximum 
above the conduction minimum of GaN.
In such a case the strained material will donate electrons to the barrier material. 
A sheet of InSb in GaN would act as a delta-doping layer. 
On InN substrates we find that the strained structures are either type II or have a broken gap.
Common to all these cases is that the compressive strain opens up the band-gap and that the top of the valence band moves up in energy compared with the unstrained situation both for compressive and  tensile strain. 
Also, the spread in energies increases with increasing strain.

\subsection{GaAs substrate}
\begin{figure}
\includegraphics[width=1.0\columnwidth]{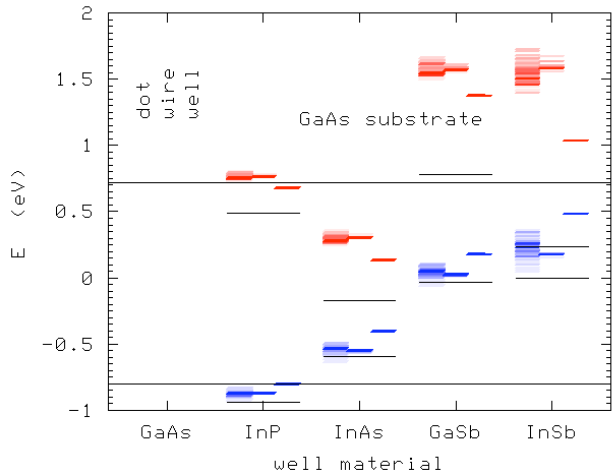}
\caption{Band edge diagram of strained dots, wires and wells on GaAs.}
\label{GaAs}
\end{figure}
Figure \ref{GaAs} shows the energies for InP, InAs, GaSb, and InSb strained to GaAs, one of the most commonly used substrates. 
We find that the band-edges of InP structures are nearly aligned with GaAs, with only small differences between InP wells, wires and dots. 
For materials with greater mismatch to GaAs than InP we find that wires and dots have similar bandedge profiles, but the gaps for  wells are substantially smaller. 
This is generally true for all non-nitride systems in which the mismatch is not too large. 
The reason is simply that wells can relax freely in the growth direction, and resulting in a lower hydrostatic strain  than for wires and dots. 
For sufficiently strained systems there is a strong interaction between the valence band and the conduction band allowing quantum wells to have a larger bandgap than dots.
For InAs dots we find an edge-to-edge gap of about 0.9 eV in agreement with previous calculations  \cite{Pryor.prb.1998, Stier.prb.1999}.
GaSb and InSb are both strongly type II for all geometries. 

Experiments on GaSb/GaAs dots show strong photoluminescence (PL) with a gap of $E_g \approx 1.1  \, \mathrm{eV}$  \cite{Bennett.jvst.1996},\cite{Hatami.apl.1995}, and pump power dependence indicating a type II structure with hole confinement.
Our calculations also indicate a type II structure with hole confinement.
Measurements of uncapped GaSb/GaAs dots show them to have a diameter  $d = 28\, \mathrm{nm}$  with a height $h=3.3  \, \mathrm{nm} $ \cite{Bennett.jvst.1996}.
To estimate the confinement energy, we approximate these dots as cylinders with the same diameter and total volume as a lens-shaped cap of the measured dimensions  ($d= 28 \, \mathrm{nm} $ and $h_{cyl} = 1.68 \, \mathrm{nm} $).  
Using the spatially averaged hole effective mass for a GaSb/GaAs dot from table III ($m_{eff} = 0.097$) and a barrier height of $0.84 \, \mathrm{eV} $, the confinement energy for the growth direction is $380  \, \mathrm{meV} $, and the confinement energy for the transverse direction is  $ 10 \, \mathrm{meV} $.
Adding the confinement energy to the (spatially averaged) edge-to-edge gap  gives $E_g =  1.06 \, \mathrm{eV} $, in excellent agreement with the measured value $1.1  \, \mathrm{eV} $.

InSb structures on GaAs are similar to GaSb on GaAs, with type II alignment and hole confinement for wells, wires and dots.  The calculated gaps are smaller however.   PL experiments on InSb/GaAs dots indicate $E_g \approx 1.1 \, \mathrm{eV}$\cite{Bennett.jvst.1996}, while our estimate of the confinement energy (using $h=5.1\, \mathrm{nm}$, $d=67\, \mathrm{nm}$) gives $E_g =0.88\, \mathrm{eV}$.  This discrepancy suggests that some alloying of the dot material has occurred or the covered dots are substantially smaller than the measurements of uncovered dots indicate. 
It is likely that the dots became smaller during capping since the GaAs was deposited using migration enhanced epitaxy\cite{Bennett.jvst.1996}.

\subsection{InP substrate}
\begin{figure}
\includegraphics[width=1.0\columnwidth]{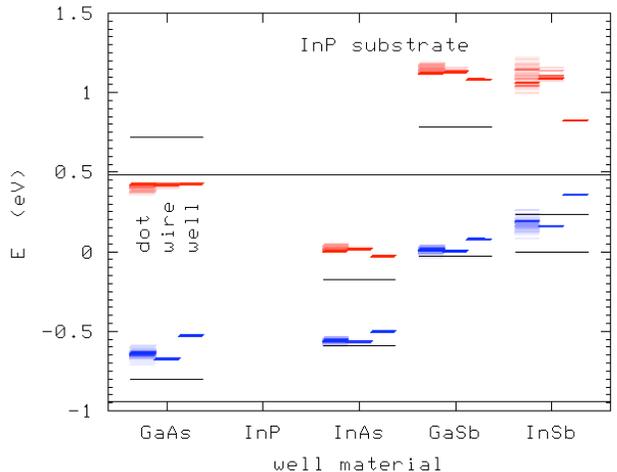}
\caption{Band edge diagram of strained dots, wires and wells on InP.}
\label{InP}
\end{figure}
Figure \ref{InP} shows the energies for materials strained to InP. 
On InP substrates we find that GaAs should be a type I quantum well with a bandgap of about 1 eV. 
Experiments on $1.8\, \mathrm{ nm}$ and $2.8\, \mathrm{nm}$ GaAs quantum wells on InP find a transition energy of  $1.148\, \mathrm{eV}$  and $1.088\, \mathrm{eV}$   respectively \cite{Pistol.prb.1992}.
Our calculations of confinement energy give transition energies of $1.234\, \mathrm{eV}$ and $1.150\, \mathrm{eV}$. 
It should be noted that the GaAs thicknesses are somewhat uncertain since they were determined from the growth, and were not directly measured\cite{Pistol.prb.1992}.  
Also, in reference \cite{Pistol.prb.1992}  the quantum wells were interpreted using calclulations that with  type II alignment and  hole confinement.  
Our calculated band alignment in Fig. \ref{InP} is nearly type II, with only a $60\, \mathrm{meV}$  barrier for the electrons. 
This discrepancy in band alignments is due to different material parameters.

InAs dots in InP have been shown to be type I and to emit light at an energy of about $0.8 \, \mathrm{eV}$
\cite{Pettersson.prb.2000}.  
The size of the dots in Reference \cite{Pettersson.prb.2000} was determined by fitting the experimental data to detailed 8-band ${\boldsymbol {k\cdot p}}$ calculations, giving dimensions of $45 \, \mathrm{nm} \times 35 \, \mathrm{nm} \times 6 \, \mathrm{nm} $.
The confinement energy is dominated by the $6 \, \mathrm{nm} $ dot height, and we obtain an estimated gap of $E_g = 0.8$, in excellent agreement with experiment.
(The confinement energy associated with the long dimensions of the dot is an order of magnitude smaller than that coming from the dot height.)
InSb quantum dots grown on InP have been found to emit photons with an energy of about 1 eV and 
were interpreted to have a type II band alignment \cite{Utzmeier.jcp.1997}.
This type II alignment has been confirmed by photoreflectance measurements on InSb islands which have been partially covered by InP giving a type I - type II transition with increasing cap layer thickness  \cite{Prieto.prl.1998}. 
The dots in Reference \cite{Utzmeier.jcp.1997} where found to be $24 \pm 4\, \mathrm{ nm}$ in diameter, and $6 \pm 3\, \mathrm{ nm}$ high, as measured by AFM on uncapped dots.
From these dimensions we obtain a confinement energy of $0.34 \, \mathrm{eV}$, giving a total gap of $E_g = 0.6\, \mathrm{eV}$.
The discrepancy between calculated and measured gaps indicates the dots probably shrunk during deposition of the InP cap layer.

\subsection{InAs substrate}
\begin{figure}
\includegraphics[width=1.0\columnwidth]{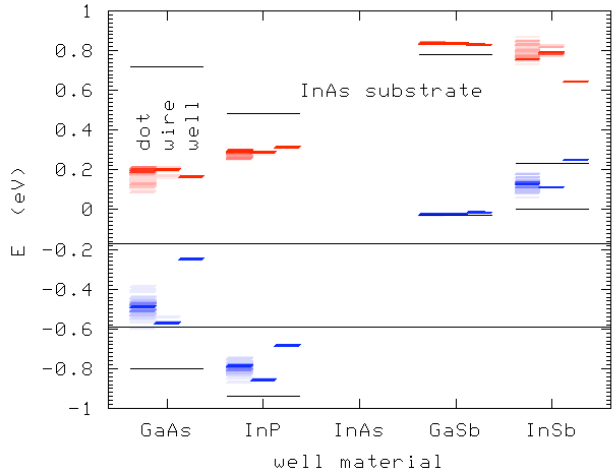}
\caption{Band edge diagram of strained dots, wires and wells on InAs.}
\label{InAs}
\end{figure}
Figure \ref{InAs} shows the energies for materials strained to InAs. 
On InAs, only GaAs has any confinement  (type II hole confinement).
On InAs substrates  both GaSb and InSb are expected to have broken gaps, with the valence band edge of GaSb and InSb above the conduction band minimum of InAs. 
This has been observed and has many interesting consequences for the electronic structure  due to charge transfer.  While broken gap superlattices have been studied
\cite{Chang.apl.1979,Altarelli.prb.1983,Krier.apl.2000,Roslund.jjap.2000}, lower dimensional structures remain unexplored.

\begin{figure}
\includegraphics[width=1.0\columnwidth]{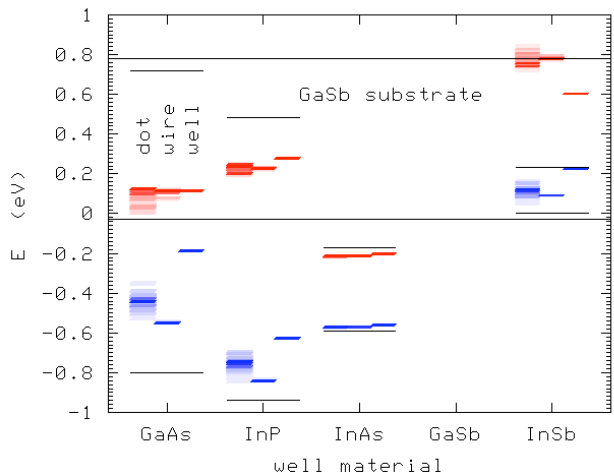}
\caption{Band edge diagram of strained dots, wires and wells on GaSb.}
\label{GaSb}
\end{figure}
\begin{figure}
\includegraphics[width=1.0\columnwidth]{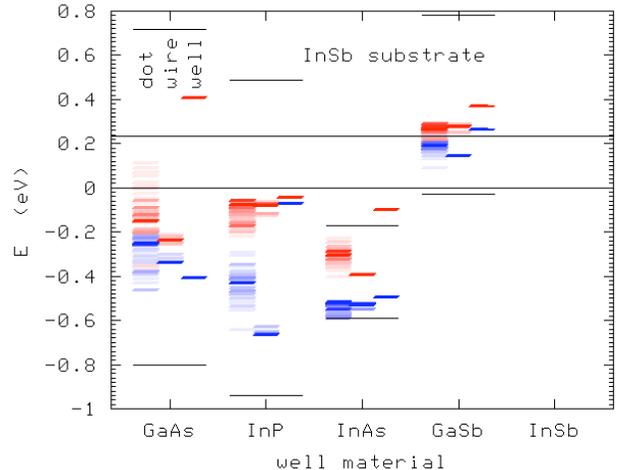}
\caption{Band edge diagram of strained dots, wires and wells on InSb.  
For all well materials the gaps are negative, with the state containing mostly valence character being higher in energy. }
\label{InSb}
\end{figure}

\begin{table}
\caption{ Conduction and valence band energies (in eV) for quantum wells strained to substrates of binary materials.
}
\begin{tabular}{l|lllllll} 
\hline \hline
& \multicolumn{7}{c|}{\emph{well material}} \\

Substrate       &\emph{GaN} &\emph{InN} &\emph{GaAs} &\emph{InP} &\emph{InAs} &\emph{GaSb} &\emph{InSb} \\
\\ \hline  &\multicolumn{7}{c}{\bf{conduction}} \\ \hline \\ 

 AlN &0.713      & -0.292        & 2.452         & 1.840         & 1.112         & 3.082         & 2.279        \\
 GaN &0.659      & -0.322        & 2.288         & 1.731         & 1.020         & 2.921         & 2.162        \\
 InN &0.444      & -0.440        & 1.635         & 1.294         & 0.653         & 2.278         & 1.695        \\
 GaAs &          &               & 0.719         & 0.682         & 0.137         & 1.375         & 1.039        \\
 InP &           &               & 0.424         & 0.485         & -0.029        & 1.085         & 0.829        \\
 InAs &          &               & 0.168         & 0.313         & -0.173        & 0.832         & 0.645        \\
 GaSb &          &               & 0.117         & 0.279         & -0.202        & 0.782         & 0.608        \\
 \hline &\multicolumn{7}{c}{\bf{valence}} \\ \hline \\
         AlN &2.644      & 2.162         & 0.212         & 0.005         & -0.198        & -0.800        & -1.249       \\
 GaN &2.640      & 2.206         & 0.268         & 0.080         & -0.141        & -0.742        & -1.178       \\
 InN &1.160      & 2.380         & 0.489         & 0.381         & 0.084         & -0.510        & -0.892       \\
 GaAs &          &               & 0.800         & 0.804         & 0.400         & -0.184        & -0.491       \\
 InP &           &               & 0.528         & 0.940         & 0.501         & -0.079        & -0.363       \\
 InAs &          &               & 0.246         & 0.682         & 0.590         & 0.012         & -0.251       \\
 GaSb &          &               & 0.188         & 0.625         & 0.558         & 0.030         & -0.228       \\
\hline \hline 
 \end{tabular} 
\end{table}
\begin{table}
\caption{ Mean conduction and valence band energies (in eV) for wires strained to substrates of binary materials.
}
\begin{tabular}{l|lllllll} 
\hline \hline
& \multicolumn{7}{c|}{\emph{wire material}} \\

Substrate       &\emph{GaN} &\emph{InN} &\emph{GaAs} &\emph{InP} &\emph{InAs} &\emph{GaSb} &\emph{InSb} \\
\\ \hline  &\multicolumn{7}{c}{\bf{conduction}} \\ \hline \\ 

 AlN &0.731      & -0.128        & 1.831         & 1.481         & 0.854         & 2.464         & 2.037        \\
 GaN &0.659      & -0.205        & 1.678         & 1.349         & 0.725         & 2.265         & 1.832        \\
 InN &0.465      & -0.440        & 1.168         & 0.912         & 0.294         & 1.590         & 1.128        \\
 GaAs &          &               & 0.719         & 0.769         & 0.309         & 1.581         & 1.600        \\
 InP &           &               & 0.419         & 0.485         & 0.020         & 1.137         & 1.109        \\
 InAs &          &               & 0.198         & 0.289         & -0.173        & 0.837         & 0.795        \\
 GaSb &          &               & 0.105         & 0.230         & -0.212        & 0.782         & 0.786        \\
 \hline &\multicolumn{7}{c}{\bf{valence}} \\ \hline \\
         AlN &2.712      & 2.404         & 0.037         & -0.116        & -0.228        & -0.837        & -1.121       \\
 GaN &2.640      & 2.377         & 0.096         & -0.051        & -0.194        & -0.796        & -1.081       \\
 InN &1.900      & 2.380         & 0.372         & 0.250         & -0.009        & -0.591        & -0.886       \\
 GaAs &          &               & 0.800         & 0.867         & 0.554         & -0.025        & -0.182       \\
 InP &           &               & 0.676         & 0.940         & 0.567         & -0.007        & -0.161       \\
 InAs &          &               & 0.568         & 0.852         & 0.590         & 0.024         & -0.114       \\
 GaSb &          &               & 0.544         & 0.835         & 0.572         & 0.030         & -0.092       \\
\hline \hline 
 \end{tabular} 
\end{table}
\begin{table}
\caption{ Mean conduction and valence band energies (in eV) for dots strained to substrates of binary materials.
}
\begin{tabular}{l|lllllll} 
\hline \hline
& \multicolumn{7}{c|}{\emph{dot material}} \\

Substrate       &\emph{GaN} &\emph{InN} &\emph{GaAs} &\emph{InP} &\emph{InAs} &\emph{GaSb} &\emph{InSb} \\
\\ \hline  &\multicolumn{7}{c}{\bf{conduction}} \\ \hline \\ 

 AlN &0.733      & -0.142        & 1.849         & 1.553         & 0.938         & 2.560         & 2.186        \\
 GaN &0.659      & -0.214        & 1.660         & 1.391         & 0.783         & 2.315         & 1.948        \\
 InN &0.448      & -0.440        & 1.079         & 0.878         & 0.279         & 1.511         & 1.139        \\
 GaAs &          &               & 0.719         & 0.768         & 0.300         & 1.578         & 1.551        \\
 InP &           &               & 0.405         & 0.485         & 0.023         & 1.146         & 1.103        \\
 InAs &          &               & 0.169         & 0.281         & -0.173        & 0.839         & 0.800        \\
 GaSb &          &               & 0.078         & 0.224         & -0.212        & 0.782         & 0.780        \\
 \hline &\multicolumn{7}{c}{\bf{valence}} \\ \hline \\
         AlN &2.744      & 2.253         & 0.613         & 0.511         & 0.290         & -0.292        & -0.621       \\
 GaN &2.640      & 2.270         & 0.639         & 0.566         & 0.321         & -0.257        & -0.563       \\
 InN &1.658      & 2.380         & 0.732         & 0.761         & 0.441         & -0.124        & -0.337       \\
 GaAs &          &               & 0.800         & 0.876         & 0.542         & -0.042        & -0.232       \\
 InP &           &               & 0.641         & 0.940         & 0.563         & -0.010        & -0.180       \\
 InAs &          &               & 0.490         & 0.794         & 0.590         & 0.023         & -0.127       \\
 GaSb &          &               & 0.443         & 0.756         & 0.568         & 0.030         & -0.114       \\
\hline \hline 
 \end{tabular} 
\end{table}
\subsection{GaSb and InSb substrates}
Figures \ref{GaSb} and \ref{InSb} shows the energies for materials strained to GaSb and InSb, respectively. 
Very little experimental work has been done on these substrates. However GaSb has a very similar lattice constant to InAs and the band edges of strained structures are thus very similar on these two substrates. 
The band alignment is very different though and we note as an example that thin layers of InAs will donate holes to GaSb. 

InSb has the interesting feature that all materials strained to it have negative gaps.
For wires and wells this inversion of the conduction and valence bands should give rise to a semi-metallic structure.
For dots there will still be discrete confined states, however the amounts of valence and conduction Bloch states will be more mixed than usual.
It should be noted that such effects are only seen in models with coupled valence and conduction bands, and would be missed in single-band models.

\begin{figure}
\includegraphics[width=1.0\columnwidth]{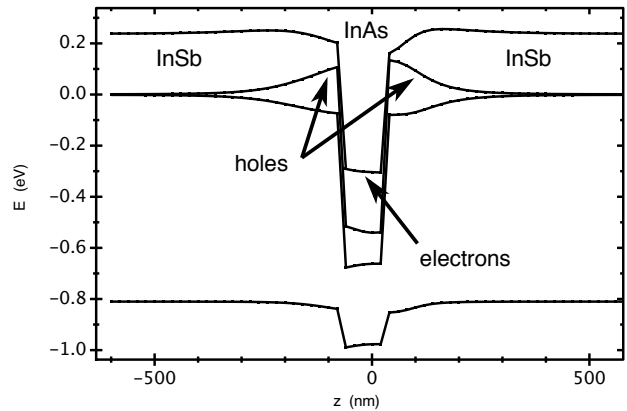}
\caption{Band diagram of an InAs/InSb dot along the [001] direction through the center of the dot.
Electrons are strongly confined in the InAs dot, while holes see a strain induced potential well in the InSb barrier adjacent to the dot.
}
\label{BD_InAs-InSb}
\end{figure}

\subsection{Broken gap structures}

For wires and dots 
the strain extends into the barrier material, affecting the electronic structure of the barrier as well.
Fig.  \ref{BD_InAs-InSb} shows an InAs dot strained to InSb, in which the InSb barrier experiences a sufficiently strong strain that the band-gap is substantially reduced near the dot.
Due to the broken gap structure the InAs acts as an "artificial acceptor", but the holes see a confining potential from the InSb  around the InAs dot.  
This results in a charged shell structure in which the InAs artificial acceptor has a negative charge which is surounded by a positive charge bound to the strained InSb.
For stacks of such dots, one would obtain sem-imetallic wires in which the core contains electrons, and the surounding shell contains holes.

\subsection{Quantum well alloys}

\begin{figure}
\includegraphics[width=1.0\columnwidth]{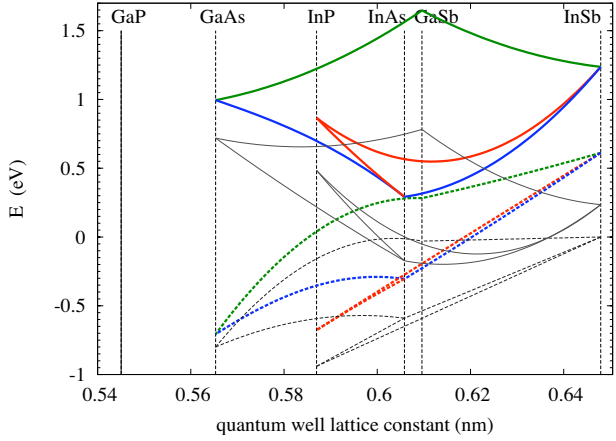}
\caption{Band edge diagram of alloyed strained wells on GaP.
The x-axis is the lattice constant of the ternary alloy comprising the quantum well.
Dotted lines are the valence band energies, solid lines are the conduction band energies, and the lighter lines are the energies for unstrained materials.
}
\label{GaPalloy}
\end{figure}

\begin{figure}
\includegraphics[width=1.0\columnwidth]{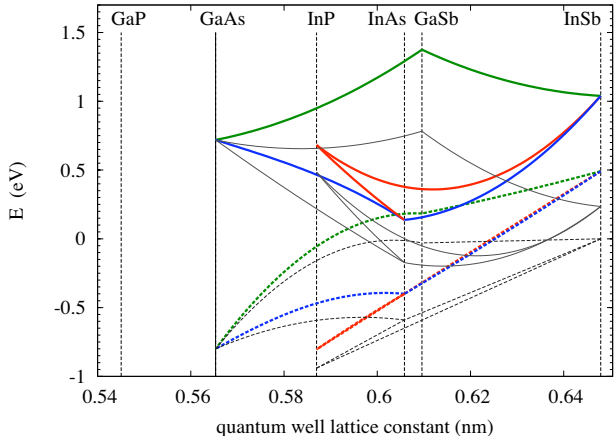}
\caption{Band edge diagram of alloyed strained wells on GaAs.
Note that when the substrate and well are lattice matched the energies cross.
}
\label{GaAsalloy}
\end{figure}

\begin{figure}
\includegraphics[width=1.0\columnwidth]{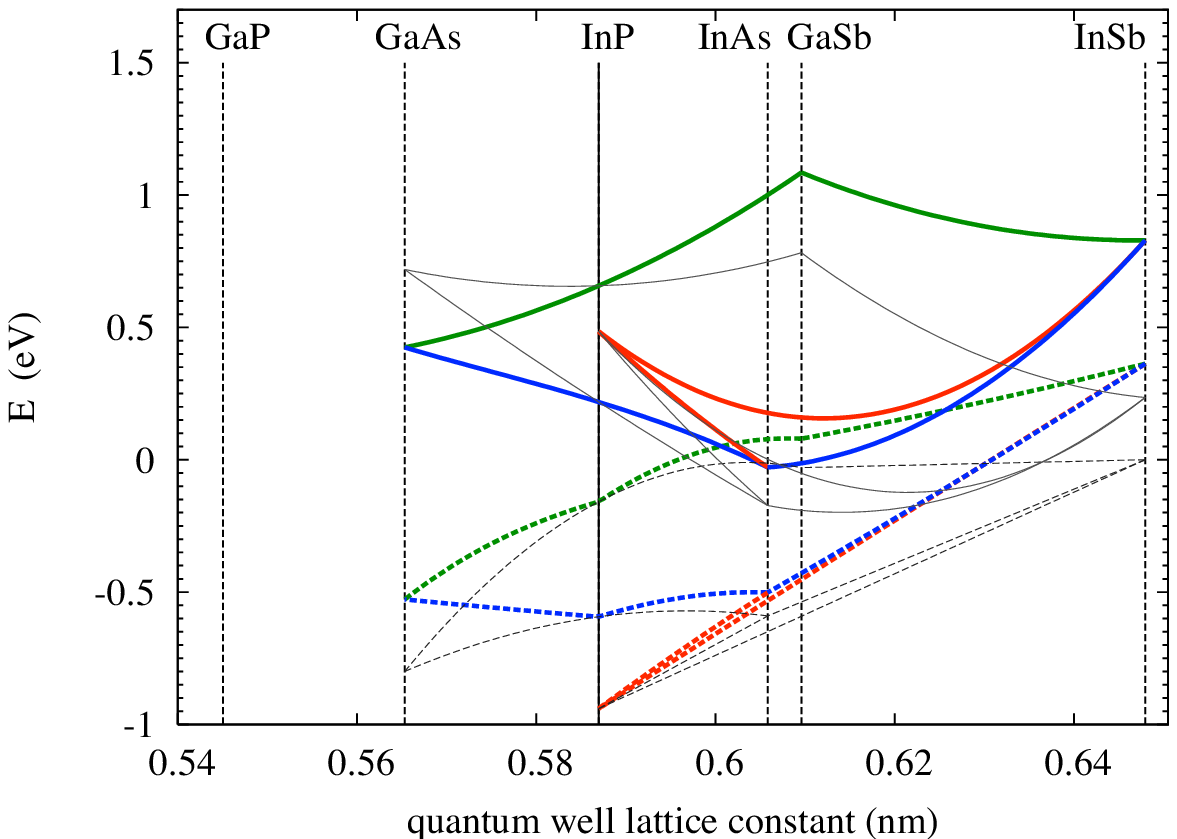}
\caption{Band edge diagram of alloyed strained wells on InP.}
\label{InPalloy}
\end{figure}

\begin{figure}
\includegraphics[width=1.0\columnwidth]{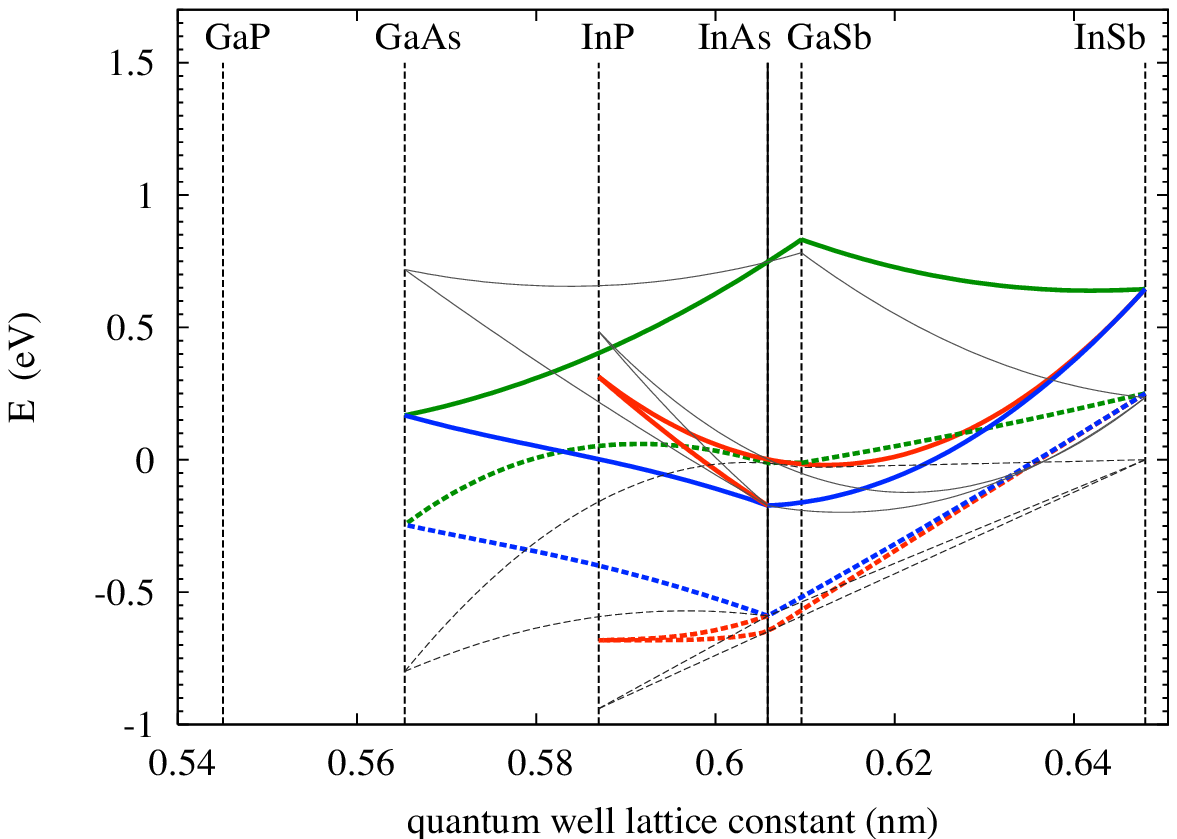}
\caption{Band edge diagram of alloyed strained wells on InAs.}
\label{InAsalloy}
\end{figure}

\begin{figure}
\includegraphics[width=1.0\columnwidth]{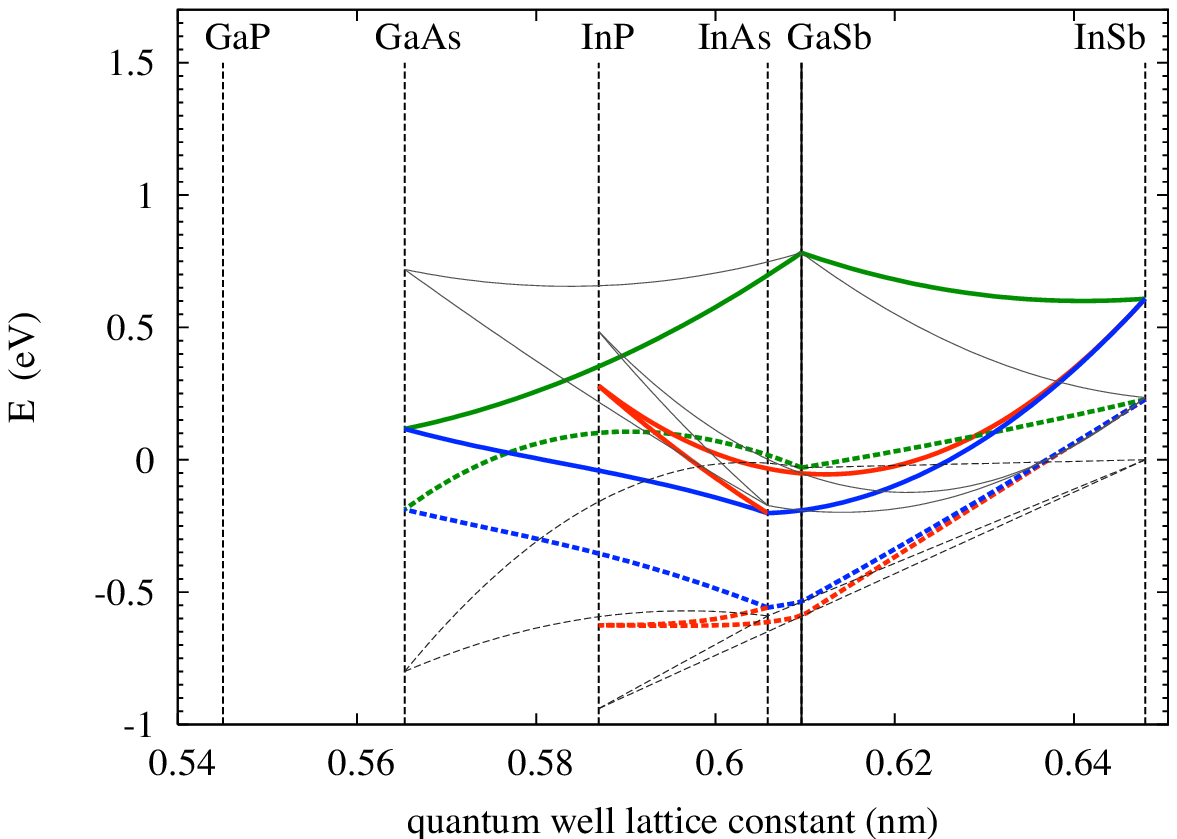}
\caption{Band edge diagram of alloyed strained wells on GaSb.}
\label{GaSballoy}
\end{figure}

\begin{figure}
\includegraphics[width=1.0\columnwidth]{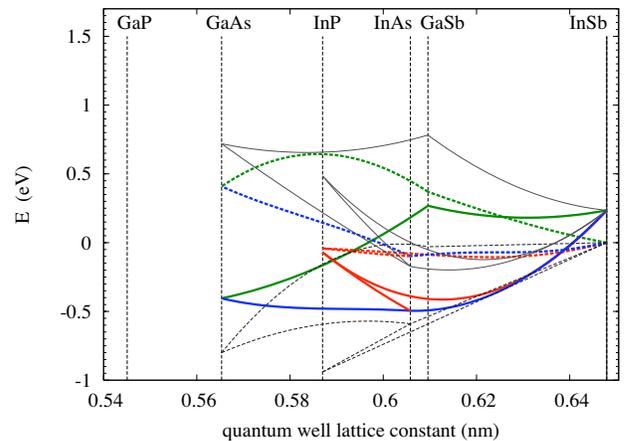}
\caption{Band edge diagram of alloyed strained wells on InSb.}
\label{InSballoy}
\end{figure}

We now turn to quantum wells composed of ternary alloys on substrates with different lattice constants.
We have calculated the band-edges of the ternary alloys interpolating among GaAs, InP, InAs, GaSb and InSb, when strained to substrates with increasing lattice constants ranging from that of GaP to InSb. 
Figures \ref{GaPalloy}-\ref{InSballoy} show the band-edges of the alloys strained to binary substrates (GaAs, InP, InAs, GaSb and InSb). 


As expected, the conduction bands decrease in energy with increasing  substrate lattice constant for all compounds.
The valence bands increase in energy for both compressive and tensile strain. 
In the absence of strain the top of the valence band consists of degenerate heavy and light hole states.
These valence band states are relatively insensitive to hydrostatic strain, but split under biaxial strain. 
Since this splitting raises one of the bands and lowers the other, the valence band edge increases regardless of the sign of the biaxial strain.
For sufficiently large substrate lattice constants the the gap may become very small and even negative (e.g. GaAs on InSb). 
We note that for substrate lattice constants $> 0.62 \, \mathrm{nm} $ we begin to see negative bandgaps, which could be useful for small-bandgap applications.
Substrates with almost arbitrary lattice constants  may be obtained using flexible substrates \cite{Eisenbeiser.ieeeedl.2002}.
Such structures could also be obtained from freestanding wires (whiskers) with properly selected alloy composition. 

A well material that is nominally metallic due to strain may obtain a gap due to confinement.
It may thus be possible to obtain narrow gap quantum wells, provided the well thickness can be made sufficiently large that the confinement energy is not too large.
With increasing lattice mismatch the bandgap of the well material decreases, but the quantum well thickness decreases as well \cite{Matthews.jcg.1974},  thus increasing the confinement energy.
Therefore designing a narrow gap quantum well  requires tradeoffs between the bandgap of the strained material and the thickness of the quantum well.

\subsection{Effective masses}
Confinement effects will increase bandgaps over the results obtained above.
It is simply impossible to cover all sizes and cases, so we have instead calculated the effective masses for electrons and holes, which can then be used to estimates the confinement energy. 
Such single-band  calculations can be quite accurate, especially for quantum wells.  
Even for quantum dots a single band approximation using a strain-dependent effective mass gives good estimates of the gap \cite{Pryor.prb.1998}.
In addition to the effective mass, the confinement energy will be effected by the geometry. 
However,  the confinement energy is primarily determined by the smallest dimension of the dot, with the detailed shape playing a smaller role \cite{Pryor.prb.1999}. 
\begin{table}
\caption{Mean effective masses of electrons and holes for strained quantum wells on different substrates.  
Values in {\it italics} are experimental electron effective masses for bulk materials.  
The calculated values differ because the 8-band model does not include the effects of remote bands.
Since strain splits the heavy-hole light-hole degeneracy, the hole masses are for the doubly degenerate highest valence state.
Hole masses for unstrained systems are excluded because of the ambiguities due to the heavy-hole light-hole degeneracy.
}
\begin{tabular}{l|lllllll} 
\hline \hline
& \multicolumn{7}{c|}{\emph{Well material}} \\

Substrate & \emph{GaN} & \emph{InN} & \emph{GaAs} & \emph{InP} &  \emph{InAs} & \emph{GaSb} & \emph{InSb} \\ \hline 
&\multicolumn{7}{c}{\bf{electrons}} \\ \hline \\
AlN  & 0.120 & 0.074 & 0.088 & 0.088 & 0.051 & 0.084 & 0.054 \\
GaN  & 0.117 & 0.074 & 0.086 & 0.088 & 0.050 & 0.082 & 0.053 \\
           & \it 0.15  &   &   &   &   &   &   \\
InN  & 0.091 & 0.072& 0.075 & 0.084 & 0.045 & 0.072 & 0.048 \\
        &             & \it 0.12 &   &   &   &   &   \\
GaAs     &   &   & 0.053     & 0.073 & 0.034 & 0.053 & 0.037 \\
               &   &   & \it 0.067 &    &   &   &   \\
InP    &   &   & 0.042 & 0.066     & 0.029 & 0.045 & 0.032 \\
          &   &   &            & \it 0.080 &   &   &   \\
InAs     &   &   & 0.028 & 0.058 &  0.023     & 0.036& 0.028 \\
             &   &   &            &             & \it 0.026  &  &   \\
GaSb     &   &   & 0.024 & 0.055   & 0.021 & 0.035     & 0.027 \\ 
               &   &   &             &              &            & \it 0.039 &              \\ \hline
&\multicolumn{7}{c}{\bf{holes}} \\ \hline \\

AlN  & 0.390 & 0.233 & 0.164 & 0.181 & 0.063 & 0.096 & 0.039 \\
GaN  &   & 0.244 & 0.164 & 0.183 & 0.063 & 0.096 & 0.039 \\
InN  & 0.267 &   & 0.159 & 0.192 & 0.061 & 0.092 & 0.038 \\
GaAs     &   &   &   & 0.207 & 0.057 & 0.083 & 0.035 \\
InP     &   &   & 0.118 &   & 0.137 & 0.078 & 0.034 \\
InAs     &   &   & 0.074& 0.194 &   & 0.074& 0.032 \\
GaSb     &   &   & 0.061& 0.183 & 0.080 &   & 0.032 \\ \hline \hline
\end{tabular}

\end{table}
\begin{table}
\caption{mean effective masses of electrons and holes for strained quantum wires on 
different substrates}
\begin{tabular}{l|lllllll} 

\hline \hline
& \multicolumn{7}{c|}{\emph{Wire material}} \\

Substrate  & \emph{GaN} & \emph{InN} & \emph{GaAs} & \emph{InP} & 
\emph{InAs} & \emph{GaSb} & \emph{InSb} \\ \hline 
&\multicolumn{7}{c}{\bf{electrons}} \\ \hline \\
AlN   & 0.122 & 0.082 & 0.078 & 0.088 & 0.051 & 0.076 & 0.062 \\
GaN   & 0.117  & 0.080 & 0.075 & 0.085 & 0.046 & 0.071 & 0.055 \\
InN   & 0.099 & 0.072  & 0.064 & 0.075 & 0.031& 0.055 & 0.028 \\
GaAs     &   &   & 0.053  & 0.076 & 0.045 & 0.061 & 0.065 \\
InP     &   &   & 0.043 &  0.066 & 0.032 & 0.047 & 0.049 \\
InAs     &   &   & 0.034 & 0.058 & 0.023  & 0.037 & 0.038 \\
GaSb     &   &   & 0.031 & 0.055 & 0.020 & 0.035  & 0.038 \\ \hline 
&\multicolumn{7}{c}{\bf{holes}} \\ \hline \\

AlN   & 0.456 & 0.354 & 0.155 & 0.193 & 0.082 & 0.101 & 0.059 \\
GaN  &         & 0.353 & 0.154 & 0.191 & 0.077 & 0.098 & 0.056 \\
InN  & 0.414 &           & 0.152 & 0.194 & 0.059 & 0.086 & 0.036 \\
GaAs     &   &   &        & 0.254 & 0.084 & 0.100 & 0.053 \\
InP     &   &   & 0.115 &           & 0.064 & 0.085 & 0.047 \\
InAs     &   &   & 0.088 & 0.184 &         & 0.071 & 0.041 \\
GaSb     &   &   & 0.078 & 0.178 & 0.046 &       & 0.040 \\ \hline \hline
\end{tabular}

\end{table}
\begin{table}
\caption{mean effective masses of electrons and holes for strained quantum dots on 
different substrates}
\begin{tabular}{l|lllllll} 

\hline \hline
& \multicolumn{7}{c|}{\emph{Dot material}} \\

Substrate  & \emph{GaN} & \emph{InN} & \emph{GaAs} & \emph{InP} &
 \emph{InAs} & \emph{GaSb} & \emph{InSb} \\ \hline 
&\multicolumn{7}{c}{\bf{electrons}} \\ \hline \\
AlN     & 0.121 & 0.080 & 0.082 & 0.095 & 0.062 & 0.082 & 0.072 \\
GaN  & 0.177   & 0.078 & 0.077 & 0.092 & 0.057 & 0.077 & 0.066 \\
InN  & 0.097 & 0.072  & 0.063 & 0.078 & 0.041 & 0.057 & 0.045 \\
GaAs     &   &     & 0.053  & 0.076 & 0.044 & 0.060 & 0.061 \\
InP     &   &   & 0.042 & 0.066  & 0.032 & 0.047 & 0.047\\
InAs     &   &   & 0.032 & 0.057 & 0.023  & 0.036 & 0.037 \\
GaSb     &   &   & 0.028 & 0.054 & 0.020 & 0.035  & 0.036 \\ \hline 
&\multicolumn{7}{c}{\bf{holes}} \\ \hline \\

AlN    & 0.441 & 0.534 & 0.189 & 0.275 & 0.085 & 0.110 & 0.049 \\
GaN  &             & 0.539 & 0.183 & 0.269 & 0.082 & 0.106 & 0.048 \\
InN    & 0.343 &               & 0.162 & 0.239 & 0.068 & 0.092 & 0.042 \\
GaAs &           &               &             & 0.237 & 0.074 & 0.097 & 0.048 \\
InP    &            &               & 0.124 &             & 0.063 & 0.085 & 0.044 \\
InAs     &         &               & 0.097 & 0.211 &             & 0.075 & 0.040 \\
GaSb   &         &              & 0.085 & 0.202 & 0.043 &              & 0.040 \\ \hline \hline
\end{tabular}

\end{table}

Effective masses were calculated by numerically computing $E({\boldsymbol k})$ at ${\boldsymbol k} = 0, \pm \delta {\boldsymbol k}$ and fitting $E({\boldsymbol k}) = {\hbar^2 \boldsymbol k}^2 / 2 m^*$  with $|\delta {\boldsymbol k}|$ equal to $10^{-2}$ of the Brillouin zone.
Anisotropy was accounted for by taking $\delta {\boldsymbol k}$ in the x-,y-, and z-directions and averaging the masses.  
The results are contained in Tables I-III which give the mean effective masses, spatially averaged over the well material, and over directions in which there is confinement.
Since the heavy-hole light-hole degeneracy is split by strain  the hole effective masses are those for the highest valence state (primarily heavy-hole).

Results for InSb substrates were omitted because all well materials have negative gaps (i.e. strain causes  the state with primarily valence character to be higher than the conduction state.)
Nitride materials on non-nitride substrates were omitted because of the extremely large range of effective masses throughout the well material, and all nitride material were assumed to be in the zincblende form.

Table I (for quantum wells) includes both the calculated and experimental electron effective masses, which differ because the 8-band model does not include the effects of remote bands.
The effective electron  mass  is given by
\begin{eqnarray}
{m^*_e} = m_0 \left[ (1+2F) + \frac{E_P(E_g+2\Delta_{SO}/3)}{E_g(E_g+\Delta_{SO})} \right]^{-1}
\end{eqnarray}

\begin{eqnarray}
F = \frac{1}{m_0} \sum_r \frac{|\langle S| p_x | u_r \rangle |^2}{E_c - E_r}
\end{eqnarray}
where $E_g$ is the gap, $\Delta_{SO}$ is the spin-orbit coupling, $E_P$ is the Kane matrix element, and $F$ is the Kane parameter for the effects of remote bands, where the index $r$ goes over the remote bands.
For most of the materials, the 8-band effective mass differs from the experimental value by $10-20\%$, which is smaller than the variation in the effective mass due to strain inhomogeneities \cite{Pryor.prb.1998}.
\section{Summary}
We have calculated band-edges for strained quantum wells, circular wires and
lens-shaped dots for a large set of III-V compounds, including alloys. 
We have also calculated the effective masses which can be used as inputs for further single band calculations to obtain the electronic structure when the absolute size of the structures is known. 
These diagrams are useful for identifying materials combinations with desired band-offsets.
We have also identified material combinations for which the embedded materials behave as artificial donors or acceptors.

\acknowledgments{
We thank N. Panev, M. K.-J. Johansson, J. Persson, and M. Miller for critical reading and enlightening comments.
This work was performed within the nanometer structure consortium in Lund and supported 
by the Swedish Foundation for Strategic Research (SSF), the Swedish Research Council (VR) and in part by the European Community's Human Potential Program under contract HPRN-CT-2002-00298, [Photon Mediated Phenomena].}


\bibliography{physRevStyleJNames,semiconductor,local}   

\end{document}